\documentclass[a4paper,12pt]{article}
\usepackage[dvips]{graphicx}

\usepackage{color}
\usepackage{epsf}
\usepackage{amssymb}
\usepackage{amsmath}

\usepackage{lscape}
\usepackage{tabularx}
\usepackage{booktabs}

\makeatletter 
\def\section{\@startsection {section}{1}{\z@}{-3.5ex plus -1ex minus -.2ex}{2.3 ex plus .2ex}{\Large\bf}}
\makeatother
\def\argmax{\mathop{\mathrm{argmax}}}

\newtheorem{problem}{Problem}
\newtheorem{evaluation}{Evaluation Procedure}

\begin{document}

\title{RNA secondary structure prediction from multi-aligned sequences%
  \footnote{An invited review manuscript 
    that will be published in a chapter of 
    the book {\bf{\em Methods in Molecular Biology}}.
Note that this version of the manuscript might be different from the published version.}}
%
\author{%
  Michiaki Hamada\,$^{1,2}$\footnote{To whom correspondence should be addressed.
    Department of Computational Biology,
    Graduate School of Frontier Sciences,
    The University of Tokyo, Tokyo, Japan.
    Tel.: +81-3-5281-5271; Fax: +81-3-5281-5331; E-mail: mhamada@k.u-tokyo.ac.jp}\\
  $^{1}$The University of Tokyo, 
  $^{2}$CBRC/AIST}
\pagestyle{plain}
\date{\today}
\maketitle
%
\begin{abstract}
  %
  It has been well accepted that the RNA secondary structures of most functional 
  non-coding RNAs (ncRNAs) are closely related to their functions and are 
  conserved during evolution.
  Hence, prediction of conserved secondary structures from evolutionarily 
  related sequences is one important task in RNA bioinformatics;
  the methods are useful not only to further functional analyses of ncRNAs
  but also to improve the accuracy of secondary structure predictions 
  and to find novel functional RNAs from the genome.
  In this review, I focus on common secondary structure prediction from a 
  given {\em aligned} RNA sequence, in which one secondary structure 
  whose length is equal to that of the input alignment is predicted.
  I systematically review and classify existing tools and algorithms for the problem,
  by utilizing the information employed in the tools and by adopting
  a unified viewpoint based on maximum expected gain (MEG) 
  estimators.
  I believe that this classification will allow 
  a deeper understanding of each tool and provide users with useful information 
  for selecting tools for common secondary structure predictions.

  {\bf key words:}
  common/consensus secondary structures; 
  comparative methods; 
  multiple sequence alignment; 
  covariation; 
  mutual information; 
  phylogenetic tree;
  energy model;
  probabilistic model;
  maximum expected gain (MEG) estimators;
  %
\end{abstract}
%
%
\section{Introduction}\label{sec:intro}

Functional non-coding RNAs (ncRNAs) play essential roles 
in various biological processes, such as transcription 
and translation regulation \cite{pmid22814587,pmid22094949,pmid21245830}.
Not only nucleotide sequences but also secondary structures 
are closely related to the functions of ncRNAs, so secondary structures 
are conserved during evolution.
Hence, 
the prediction of these conserved secondary structures 
(called `common secondary structure prediction' throughout this review) from 
evolutionarily related RNA sequences  
is among the most important tasks in RNA bioinformatics, 
because it provides
useful information for further functional analysis of the targeted RNAs
\cite{pmid16495232,pmid22965121,pmid23172290,pmid23172291,Meer20032012}. 
It should be emphasized that common secondary structure predictions are 
also useful to improve the accuracy of secondary structure prediction~\cite{pmid23435231,pmid15458580}  
or to find functional RNAs from the genome \cite{pmid16273071,pmid19908359}.

\begin{figure}[th]
  \centerline{
    \begin{tabular}{cc}
      \multicolumn{1}{l}{\small {\bf(a)} Secondary structure prediction} &  
      \multicolumn{1}{l}{\small {\bf(b)} Common secondary structure prediction}\\
      \includegraphics[width=0.5\textwidth]{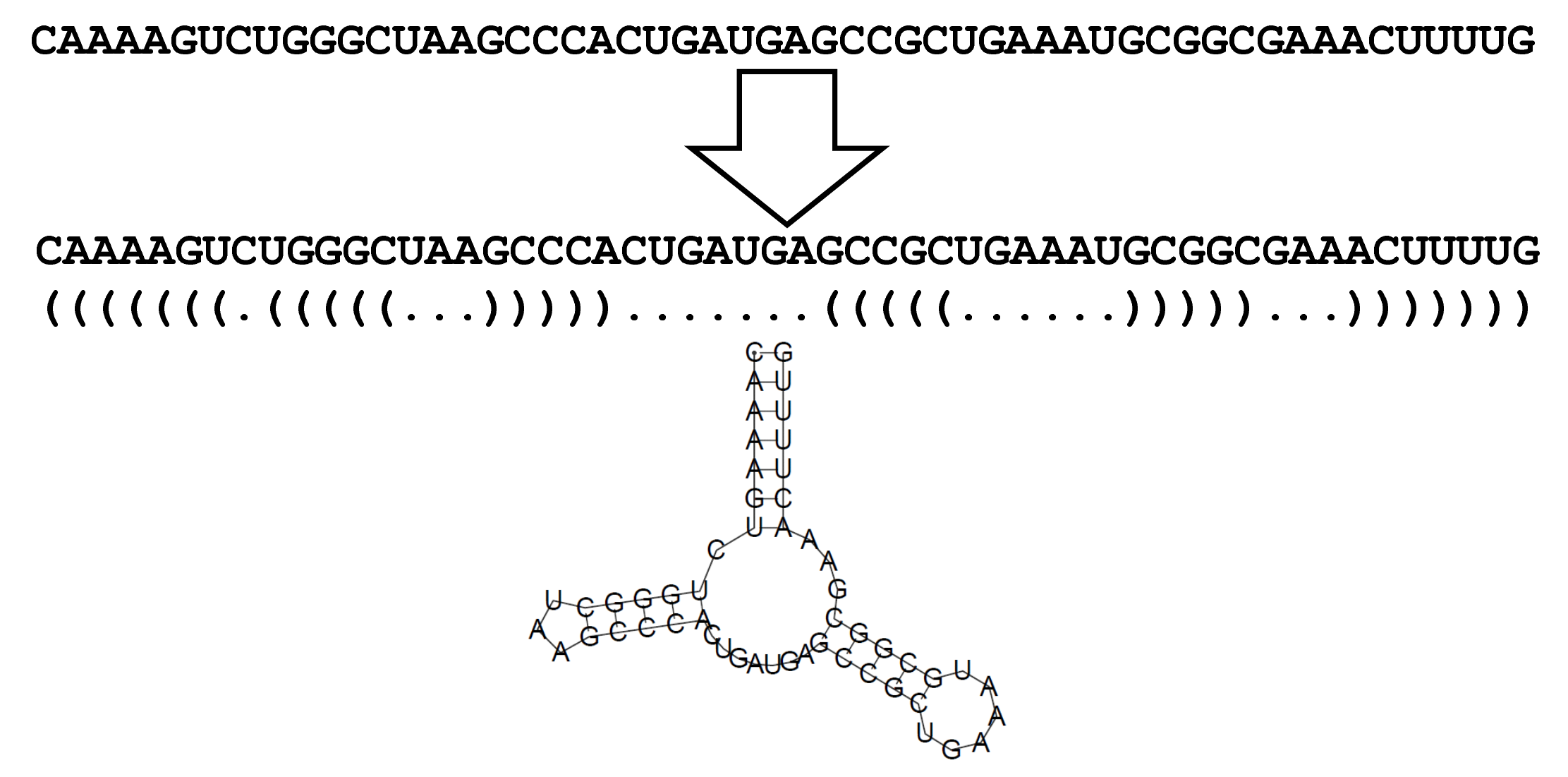} &
      \includegraphics[width=0.5\textwidth]{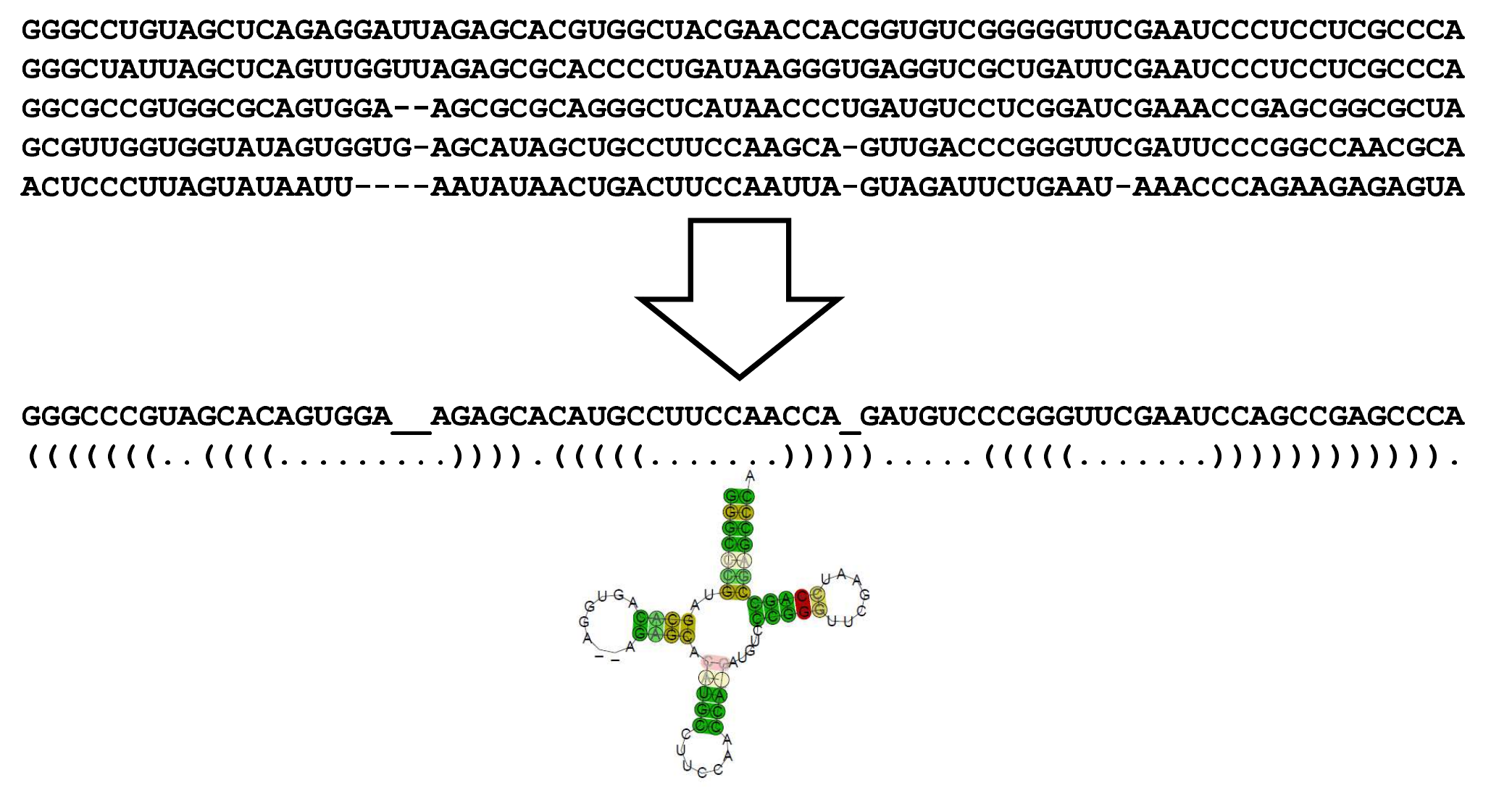} \\
    \end{tabular}
  }
  \caption{\label{fig:rna_sec_csec}\small
    {\bf (a)} Conventional RNA secondary structure prediction, 
    in which the input is an individual RNA sequence and 
    the output is an RNA secondary structure of the sequence.
    {\bf (b)} Common (or consensus) RNA secondary structure prediction in which 
    the input is a multiple sequence alignment of RNA sequences and the output
    is an RNA secondary structure whose length is equal to the length of 
    the alignment.
    The secondary structure is a called the common (or consensus) secondary 
    structure.
  }
\end{figure}

Approaches to common secondary structure prediction are divided 
into two categories with respect to the input:
(i) {\em unaligned} RNA sequences and
(ii) {\em aligned} RNA sequences. 
Common secondary structure predictions from {\em unaligned} RNA sequences 
can be solved by utilizing the Sankoff algorithm~\cite{sankoff1985}. 
However, it is known that the Sankoff algorithm has huge computational costs: 
$O(L^{3n})$ and $O(L^{2n})$ for time and space, respectively, where 
$L$ is the length of RNA sequences and $n$ is the number of input sequences.
For this reason, the second approach, whose input is {\em aligned} RNA sequences, 
is often used in actual analyses.
The problem focused on this review is formulated as follows.
%
\begin{problem}[RNA common secondary structure prediction of aligned sequences]\label{prob:rnac}
  Given an input multiple sequence alignment $A$, predict 
  a secondary structure $y$
  whose length is equal to the length of the input alignment.
  The secondary structure $y$ is called the common (or consensus) secondary 
  structure of the multiple alignment $A$.
\end{problem}
%
In contrast to conventional RNA secondary structure 
prediction (Figure~\ref{fig:rna_sec_csec}a),
the input of common secondary structure prediction 
is a multiple sequence alignment of RNA sequences 
(Figure~\ref{fig:rna_sec_csec}b) and
the output is a secondary structure with the same length as the alignment.
In general, the predicted common RNA secondary structure is expected to 
represent a secondary structure that commonly appears in the 
input alignments or is conserved during evolution.

\begin{figure}[th]
  \centerline{%
    \includegraphics[width=0.99\textwidth]{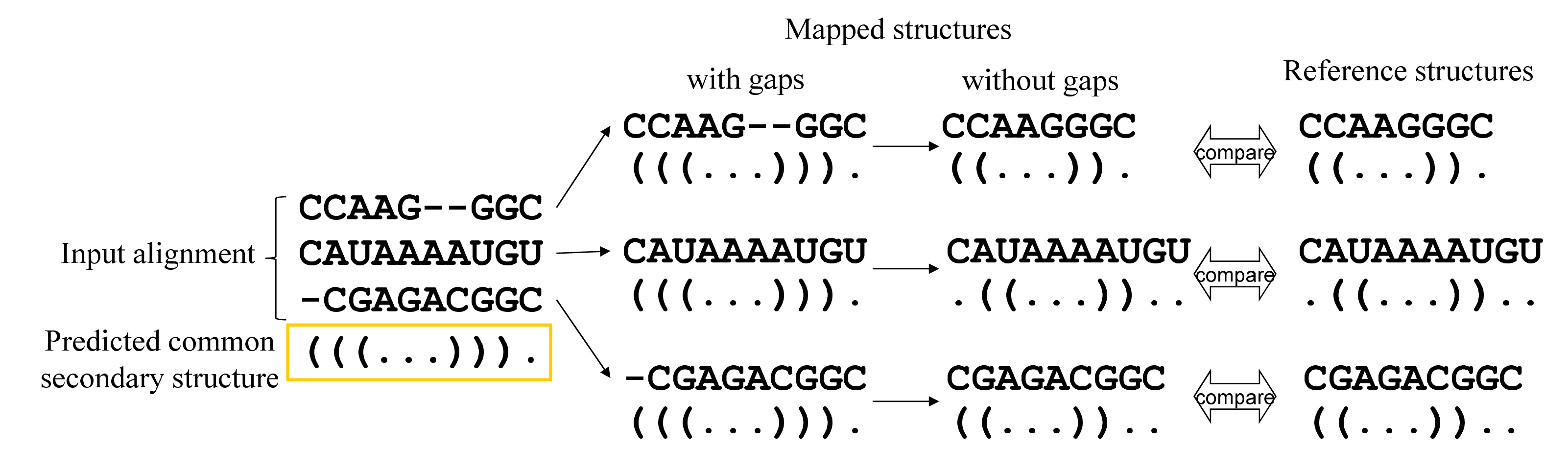}%
  }
  \caption{\label{fig:evaluation_process}\small
    An evaluation procedure for the predicted common RNA secondary structure 
    of an input alignment, in which the reference secondary structure of 
    each RNA sequence in the alignment is given.
    This procedure is based on the idea that a common secondary structure should reflect as many of
    the secondary structures of each RNA sequence in the input alignment 
    as possible.
    Mapped RNA secondary structures without gaps are computed by 
    getting rid of base-pairs that correspond to gaps.
    Note that it is difficult to compare a predicted common secondary structure 
    with a {\em reference} common secondary structure, 
    because the reference common RNA secondary structure 
    for an arbitrary input alignment is not available in general. 
    In most studies of common secondary structure prediction, evaluation is conducted
    by using this procedure or a variant.
    See \cite{pmid20843778} for a more detailed discussion of 
    evaluation procedures for Problem~\ref{prob:rnac}.
  }
\end{figure}

To develop tools (or algorithms) for solving this problem, 
it should be made clear what a {\em better} common secondary 
structure is.
However, the evaluation method for predicted 
common RNA secondary structures is {\em not} trivial.
A predicted common secondary structure depends on not 
only RNA sequences in the alignment 
but also multiple alignments of input sequences, and in general,
no reference (correct) common secondary structures 
are available\footnote{Reference common secondary 
structures are available for only reference multiple sequence alignments in
the Rfam database \cite{pmid23125362} ({http://rfam.sanger.ac.uk/})}.
A predicted common secondary structure is therefore evaluated, 
based on reference RNA secondary structures 
for each {\em individual} RNA sequence in the alignment as follows. 
%
\begin{evaluation}[for Problem~\ref{prob:rnac}]
  \label{def:eval_proc}
  Given reference secondary structures for each RNA sequence,
  evaluate the predicted common secondary structure $y$ as follows. 
  First, map $y$ onto each RNA sequence in the alignment $A$%
  \footnote{See the column `Mapped structure with gaps' 
    in Figure~\ref{fig:evaluation_process}, for example.}. 
  Second, remove all gaps in each sequence and the corresponding base pairs in the mapped secondary
  structure in order to maintain the consistency of the secondary structures%
  \footnote{See the column `Mapped structure without gaps' 
    in Figure~\ref{fig:evaluation_process}, for example.}. 
  Third, calculate the quantities TP, TN, FP, and FN for each mapped secondary structure $y^{(map)}$ 
  with respect to the reference
  secondary structure\footnote{TP, TN, FP, and FN are the respective numbers of true positive, true-negative, false-positive 
    and false-negative base-pairs of a predicted secondary structure with respect to the reference structures.}.
  Finally, calculate the evaluation measures sensitivity (SEN), positive predictive value (PPV), and Matthew correlation coefficient (MCC) for the sum of TP, TN, FP, and FN over 
  all the RNA sequences in the alignment $A$.
\end{evaluation}
%
Figure~\ref{fig:evaluation_process} shows an 
illustrative example of Evaluation Procedure~\ref{def:eval_proc}.
Note that there exist a few variants of this procedure 
(e.g., \cite{pmid19014431}).

%
%

In this review, I aim to classify the existing tools (or algorithms) for Problem~\ref{prob:rnac};
These tools are summarized in Table~\ref{tab:list_of_tools}, 
which includes all the tools for common secondary
structure prediction as of 17th June, 2013.
To achieve this aim,
I describe the information that is often utilized 
in common secondary structure predictions, and classify tools from
a unified viewpoint based on maximum expected gain  (MEG)
estimators. 
I also explain the relations between the MEG estimators 
and Evaluation Procedure~\ref{def:eval_proc}.

This review is organized as follows.
In Section~\ref{sec:info}, I summarize the information 
that is commonly utilized when designing algorithms for common secondary 
structure prediction.
In Section~\ref{sec:classification}, several concepts to be utilized 
in the classification of tools are presented, and the currently available tools are classified within this framework in Section~\ref{sec:tools}.
In Section~\ref{sec:discussion}, I discuss several points arising from this classification framework. 
Finally, conclusions are given in Section~\ref{sec:conclusion}.

\begin{table}[h]
  \caption{\label{tab:list_of_tools}
    List of tools for common secondary structure prediction 
    from {\em aligned} sequences
  }
  \centerline{
    {\small
      \begin{tabularx}{\textwidth}{llX}
  \toprule
  Tool & Ref. & Description\\
  \midrule
  \multicolumn{2}{l}{\bf(Without pseudoknot)} \\
  CentroidAlifold & \cite{pmid19095700,pmid20843778} & 
  Achieved superior performance in a recent benchmark (CompaRNA) \cite{pmid23435231}.
  Using an MEG estimator with the $\gamma$-centroid-type gain function 
  (Section~\ref{sec:gcent-gain}) 
  in combination with a mixture probability distribution,
  including several types of information (Section~\ref{sec:mix_dist})\\
  ConStruct & \cite{pmid10518612,pmid18442401} & 
  A semi-automatic, graphical tool  based on  mutual information (Section~\ref{sec:MI})\\
  KNetFold & \cite{pmid16495232} &
  Computes a consensus RNA secondary structure from an RNA sequence alignment based 
  on machine learning (Bayesian networks).\\
  McCaskill-MEA & \cite{pmid17182698} &
  Adopting majority rule with the McCaskill energy model \cite{pmid1695107}, leading to an algorithm that is robust
  to alignment errors.\\
  PETfold  & \cite{pmid18836192} & 
  Considers both phylogenetic information and thermodynamic stability 
  by extending Pfold, in combination with an MEA estimation.\\
  Pfold  & \cite{pmid10383470,pmid12824339} &  
  Uses phylogenetic tree information with simple  SCFG\\
  PhyloRNAalifold & \cite{pmid23621982} & 
  Incorporates the number of co-varying mutations on the phylogenetic tree of the aligned
  sequences into the covariance scoring of RNAalifold.\\
  PPfold  & \cite{pmid22877864} &
  A multi-threaded implementation of the Pfold algorithm, 
  which is extended to evolutionary analysis with a flexible 
  probabilistic model for incorporating auxiliary data, 
  such as data from structure probing experiments. \\
  RNAalifold & \cite{pmid19014431,pmid17993696,pmid12079347} &
  Considers both thermodynamic stability and co-variation in combination with 
  RIBOSUM-like scoring matrices \cite{pmid14499004}.\\
  RSpredict & \cite{pmid20140072} & 
  Takes into account sequence covariation and employs effective
  heuristics for accuracy improvement.\\
  \multicolumn{2}{l}{\bf (With pseudoknot)}\\
  hxmatch & \cite{pmid17048382} &
  Computes consensus structures including pseudoknots based on alignments of a few sequences. 
  The algorithm combines thermodynamic and covariation information to assign scores to all possible base pairs, the base
  pairs are chosen with the help of the maximum weighted matching algorithm\\
  ILM & \cite{pmid14693809} &
  Uses mutual Information and helix plot in combination with heuristic optimization.\\
  IPKnot & \cite{pmid21685106} &
  Uses MEG estimators with $\gamma$-centroid gains and heuristic probability distribution of
  RNA interactions together with integer linear programming to compute a decoded RNA secondary structure\\
  MIfold & \cite{pmid16000013} &
  A MATLAB(R) toolbox that employs mutual information, 
  or a related covariation measure, to display and predict 
  common RNA secondary structure\\
  \bottomrule
\end{tabularx}

  }}{\small
    To the best of my knowledge, this is a complete list of tools for the problem as of 17 June 2013.
    Note that tools for common secondary structure prediction from {\em unaligned} RNA sequences 
    are not included in this list.
    See Table~\ref{tab:comparison} for further details of the listed tools.
  }
\end{table}

\section{Common information utilized in 
common secondary structure predictions}\label{sec:info}
%
Several pieces of information are generally utilized in tools and algorithms 
for predicting common RNA secondary structures. These will now be briefly summarized.

\subsection{Fitness to each sequence in the input alignment}
%
The common RNA secondary structure should be a representative secondary
structure among RNA sequences in the alignment.  
Therefore, the fitness of a predicted common secondary structure to 
{\em each} RNA sequence in the alignment is useful information.
In particular, in Evaluation Procedure~\ref{def:eval_proc}, 
the fitness of a predicted common secondary structure to each RNA sequence is evaluated. 

This fitness is based on probabilistic models for RNA secondary structures 
of individual RNA sequence,
such as the energy-based and machine learning models 
shown in  Sections~\ref{sec:TS} and ~\ref{sec:ML}, respectively. These models provide a probability distribution of secondary structures
of a given RNA sequence\footnote{%
$p(\theta|x)$ denotes a probability distribution of RNA secondary structures 
  for a given RNA sequence $x$.
}

\subsubsection{Thermodynamic stability --- energy-based models}\label{sec:TS}

Turner's energy model \cite{pmid10329189} is an energy-based model, which
considers the thermodynamic stability of RNA secondary structures. 
This model is widely utilized
in RNA secondary structure predictions, in which 
experimentally determined energy parameters \cite{pmid10329189,pmid15123812,pmid9778347} are employed.
In the model, structures with a lower free energy are more stable than those 
with a higher free energy.
Note that Turner's energy model leads to a probabilistic model 
for RNA sequences, providing a probability distribution of secondary structures, 
which is called the McCaskill model \cite{pmid1695107}.

\subsubsection{Machine learning (ML) models}\label{sec:ML}

In addition to the energy-based models described in the previous subsection, probabilistic models for 
RNA secondary structures based on machine learning (ML) approaches have been proposed.
In contrast to the energy-based models, machine learning models can 
automatically learn parameters from training data (i.e., a set of RNA sequences
with secondary structures).
There are several models based on machine learning which adopt different approaches:
(i) Stochastic context free grammar (SCFG) models \cite{pmid15180907};
(ii) the CONTRAfold model \cite{pmid16873527} (a conditional random field model); 
(iii) the Boltzmann Likelihood (BL) model \cite{pmid20940338,pmid19933322,pmid17646296};
and
(iv) non-parametric Bayesian models \cite{DBLP:journals/jbcb/SatoHMAS10}.

See Rivas {\it et al.} \cite{pmid22194308} for detailed comparisons of probabilistic models
for RNA secondary structures.

\subsubsection{Experimental information}\label{sec:EI}

Recently, 
experimental techniques to probe RNA structure by high-throughput sequencing 
(SHAPE \cite{pmid20554050}; PARS \cite{pmid20811459}; FragSeq \cite{pmid21057495})
has enabled genome-wide measurements of RNA structure. 
Those experimental techniques stochastically estimate 
the flexibility of an RNA strand, which can be considered as a kind of 
{\em loop probability} for every nucleotide in an RNA sequence.
Remarkably, secondary structures of long RNA sequences, such as HIV-1 \cite{pmid19661910}, 
HCV (hepatitis C virus) \cite{pmid21965531}, 
and large intergenic ncRNA (the steroid receptor RNA activator) \cite{pmid22362738},
have been recently determined by combining those experimental techniques 
with computational approaches.
If available, such experimental information is useful in common secondary 
structure predictions, because they provide reliable
secondary structures for each RNA sequence.

\subsection{Mutual information}\label{sec:MI}

The mutual information of the $i$-th and $j$-th columns 
in the input alignment is defined by 
%
\begin{align}
  M_{ij} = \sum_{X,Y} f_{ij}(XY) \log \frac{f_{ij}(XY)}{f_i(X)f_j(Y)}
  =KL(f_{ij}(XY)||f_i(X)f_j(Y))\label{eq:MI}
\end{align}
where
$f_i(X)$     is the frequency of base    $X$ at alignment position $i$;
$f_{ij}(XY)$ is the joint frequency of finding $X$ in the $i$-the column and $Y$ 
in the $j$-th column, and
$KL(\cdot||\cdot)$ denotes the Kullback-Liebler distance between two probability
distributions.
As a result, the complete set of mutual information can be represented as an upper triangular matrix:
$\{M_{ij}\}_{i<j}$. 

Note that the mutual information score makes no use of base-pairing rules of RNA secondary structure.
In particular, mutual information does not account for consistent non-compensatory mutations at all,
although  information about them would be useful when predicting common secondary structures as
described in the next subsection.

\subsection{Sequence covariation of base-pairs}\label{sec:CV}

\begin{figure}
\begin{tabular}{cc}
\multicolumn{1}{l}{\bf\small(a)} & \multicolumn{1}{l}{\bf\small(b)}\\
\includegraphics[width=0.73\textwidth]{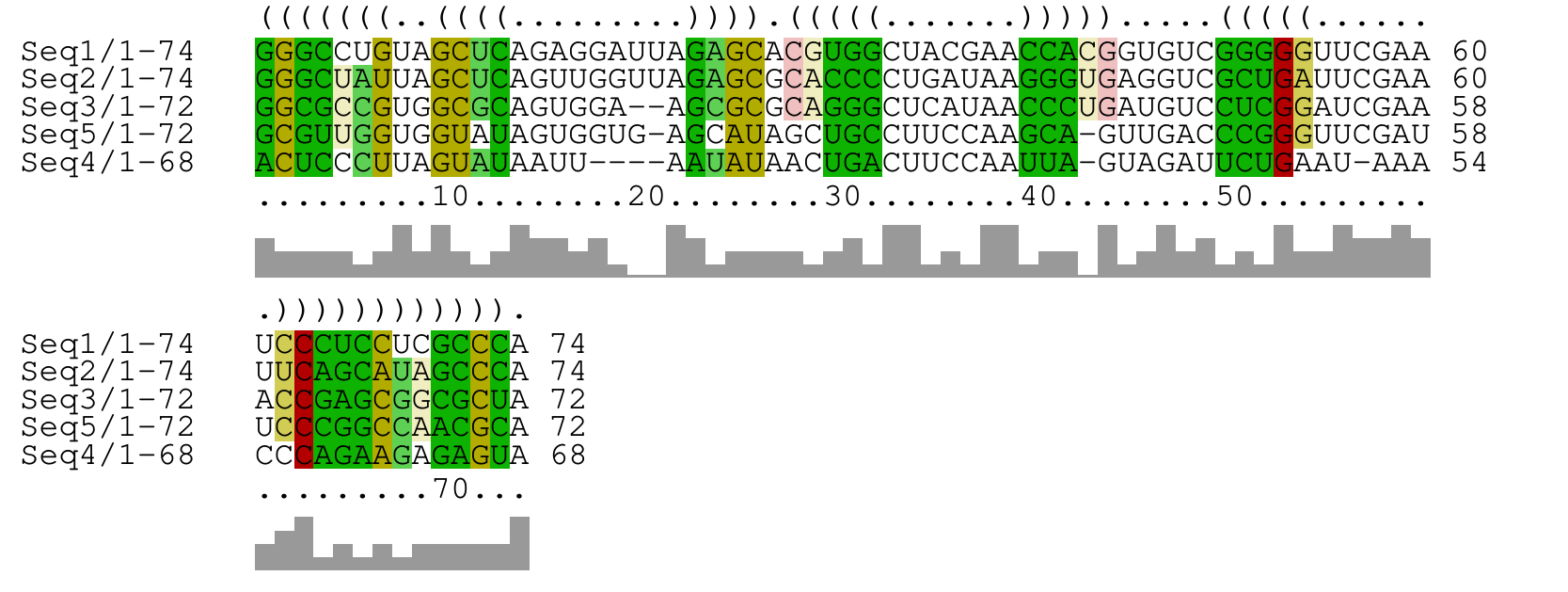} &
\includegraphics[width=0.27\textwidth]{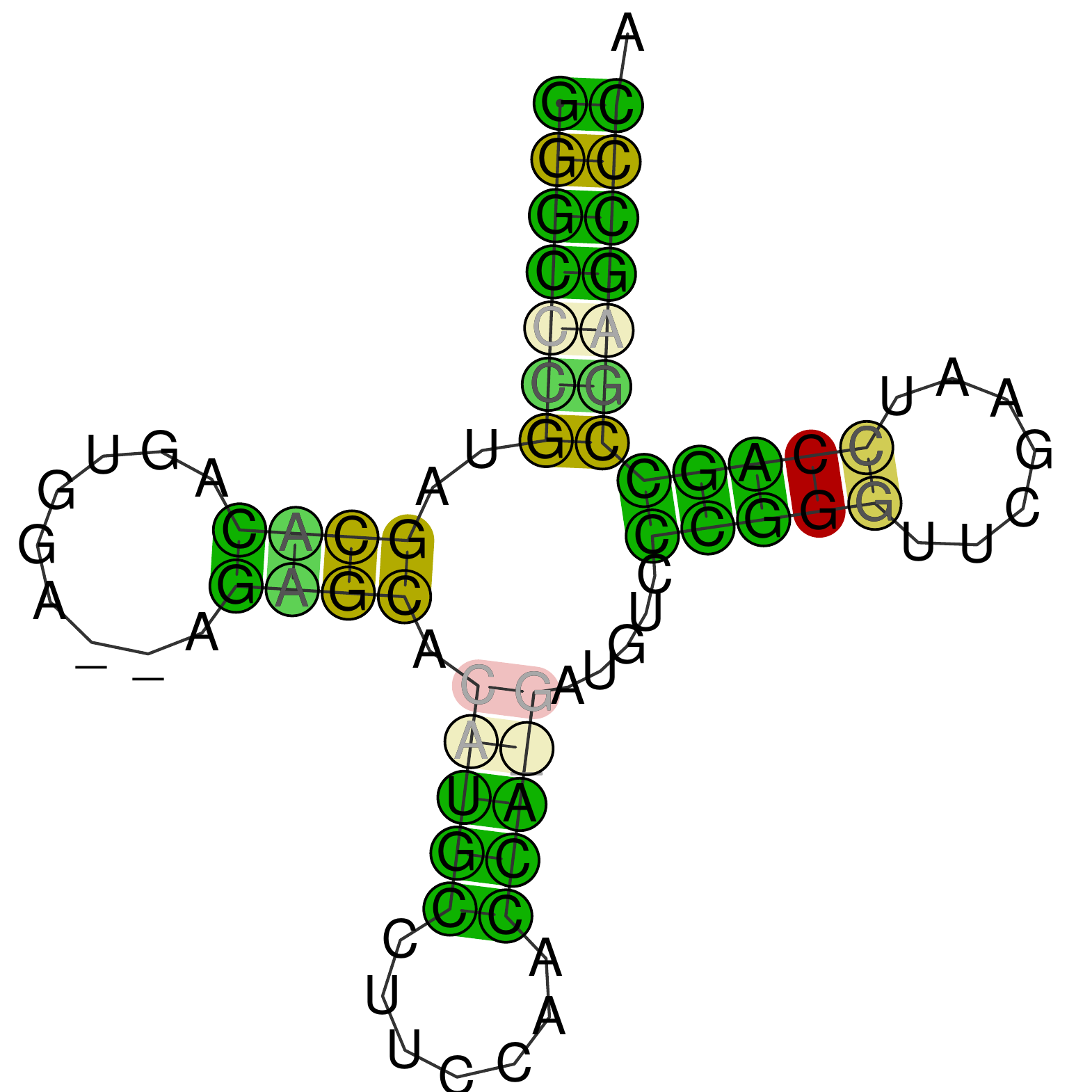} 
\end{tabular}
\caption{\label{fig:eg_cv}\small
An example of covariation of base-pairs in an alignment of tRNA sequences:
{\bf(a)} a multiple alignment of tRNA sequences and 
{\bf(b)} a predicted common secondary structure
of the alignment.
The figures are taken from the output of an example on the 
RNAalifold~\cite{pmid19014431} Web Server ({http://rna.tbi.univie.ac.at/cgi-bin/RNAalifold.cgi}).
}
\end{figure}

Because secondary structures of ncRNAs are related to their functions, 
mutations that preserve base-pairs (i.e., covariations of a base-pair) often occur
during evolution.
Figure~\ref{fig:eg_cv} shows an example of covariation of base-pairs of 
tRNA sequences, in which many covariations of base-pairs are found, especially
in the stem parts in the tRNA structure. 

The covariation of the $i$-th and $j$-th columns in the input alignment is 
evaluated by the averaged number of compensatory mutations defined by
%
\begin{align}
  C_{ij} =  \frac{2}{N(N-1)} \sum_{(x,y) \in A} d_{ij}^{x,y} \Pi_{ij}^x \Pi_{ij}^y
\end{align}
%
where $N$ is the number of sequences in the input alignment $A$. For an RNA sequence 
$x$ in $A$, $\Pi_{ij}^x=1$ if
$x_i$ and $x_j$ form a base-pair and $\Pi_{ij}^x=0$ otherwise, and
\begin{align}
  d_{ij}^{x,y}=2-\delta(x_i,y_i)-\delta(x_j,y_j)
\end{align}
where $\delta$ is the delta function: 
$\delta(a,b)=1$ only if $a=b$, and $\delta(a,b)=0$ otherwise.

For instance,
RNAalifold \cite{pmid15313604} uses the information of covariation in combination 
with the thermodynamic stability of common secondary structures.

\subsection{Phylogenetic (evolutionary) information}\label{sec:PI}

Because most secondary structures of functional ncRNAs are conserved
in evolution, the phylogenetic (evolutionary) information 
with respect to the input alignment is useful for predicting common secondary 
structures, and is employed by  several tools.
Pfold \cite{pmid10383470,pmid12824339} incorporates 
this information into probabilistic model for common secondary structures 
(Section~\ref{sec:pfold-model})
for the first time.

\subsection{Majority rule of base-pairs}\label{sec:MR}

Kiryu {\it et al.} \cite{pmid17182698} proposed the use of the majority rule of base-pairs 
in predictions of common secondary structures. This rule states that
base-pairs supported by many RNA sequences should be included in a
predicted common secondary structure.
Specifically,
Kiryu {\it et al.} utilized an averaged probability distribution of secondary structures among
RNA sequences to
predict a common RNA secondary structure from a given alignment (see Section~\ref{sec:ave-dist}).

The aim of this approach 
is to mitigate alignment errors, because the effects of 
a minor alignment errors can be disregarded in the prediction of common secondary structures.

\section{Common secondary structure predictions with MEG estimators}\label{sec:classification}

\subsection{ MEG estimators}\label{sec:MEG}

In this section, I classify the existing algorithms 
for common RNA secondary structure prediction (shown in Table~\ref{tab:list_of_tools}) from a unified viewpoint, 
based on a previous study \cite{pmid20843778}
in which the following type of estimator \cite{pmid21365017,pmid22313125} was employed.
%
\begin{align}
  \hat y=\argmax_{y \in \mathcal{S}(A)} \sum_{\theta\in Y} G(\theta,y) p(\theta|A)\label{eq:MEG}
\end{align}
where
$\mathcal{S}(A)$ denotes a set of possible secondary structures with length 
$|A|$ (the length of the alignment $A$),
$G(\theta,y)$ is called a `gain function' and returns a measure of the similarity between two common secondary structures, 
and
$p(\theta|A)$ is a probabilistic distribution on $\mathcal{S}(A)$.
This type of estimator is an MEG estimator;
When the gain function is designed according to accuracy measures 
for target problems, the MEG estimator is often
called a `maximum expected accuracy (MEA) estimator' \cite{pmid22313125}\footnote{%
The gain $G(\theta,y)$ is equal to the accuracy measure $Acc(\theta,y)$ for a prediction and 
references, so the MEG estimator maximizes the expected accuracy under a given probabilistic 
distribution.
}.

In the following, a common secondary structure $\theta\in\mathcal{S}(A)$ 
is represented
as a upper triangular matrix $\theta=\left\{\theta_{ij}\right\}_{1 \le i < j \le |A|}$.
In this matrix $\theta_{ij}=1$ if the $i$-th column in 
$A$ forms a base-pair
with the $j$-th column in $A$, and $\theta_{ij}=0$ otherwise.

The choices of $p(\theta|A)$ and $G(\theta,y)$ are described 
in Sections~\ref{sec:choice_prob} and~\ref{sec:choice_gain}, respectively.

\subsection{Choice of probabilistic models $p(\theta|A)$}\label{sec:choice_prob}

The probabilities $p(\theta|A)$ provide a distribution of common RNA secondary structures
given a multiple sequence alignment $A$. This distribution is given by the following models.

\subsubsection{RNAalipffold model}\label{sec:rnaalipffold_model}
%
The RNAalipffold model is a probabilistic version of RNAalifold \cite{pmid12079347}, 
which provides a probability distribution of common RNA secondary structures 
given a multiple sequence alignment.
The distribution on $\mathcal{S}(A)$ is defined by
\begin{align}
  p^{\mbox{\scriptsize (RNAalipffold)}}(\theta|A) 
  = \frac{1}{Z(T)}\exp\left(\frac{-E(\theta,A)}{kT} + Cov (\theta, A) \right)\label{eq:alipffold}
\end{align}
where $E(\theta,A)$ is the averaged free energy of RNA sequences in the alignment
with respect to the common secondary structure $\theta$ in $A$ 
and $Cov(\theta,A)$ 
is the base covariation (cf. Section~\ref{sec:CV}) 
with respect to the common secondary structure $\theta$.
The ML estimate of this distribution is equivalent to the prediction of RNAalifold.

Note that 
the negative part of the exponent in Eq~(\ref{eq:alipffold}) 
is called the {\em pseudo} energy and it plays an essential role 
in finding ncRNAs from multiple alignments 
\cite{pmid19908359,pmid15665081}.

\subsubsection{Pfold model}\label{sec:pfold-model}

The Pfold model \cite{pmid12824339,pmid10383470} incorporates phylogenetic (evolutionary) 
information about the input alignments into a probabilistic distribution of common 
secondary structures:
%
\begin{align}
  p^{(\mbox{\scriptsize pfold})}(\theta|A) = 
  p^{(\mbox{\scriptsize pfold})}(\theta|A,T,M)=\frac{p(A|\theta,T)p(\theta|M)}{p(A|T,M)}
\end{align}
where $T$ is a phylogenetic tree, $A$ is the input data (i.e., an alignment), $M$ is a prior model 
for secondary structures (based on SCFGs\footnote{%
  The SCFG is based on the rules $S\to LS|L$, $F\to dFd|LS$, $L\to s|dFd$.
}).
Unless the original phylogenetic tree $T$ is obtained, $T$ is taken to be the ML estimate of 
the tree, $T^{ML}$, given the model $M$ and the alignment $A$.

\subsubsection{Averaged probability distribution of each RNA sequence}\label{sec:ave-dist}

Predictions based on RNAalipffold and Pfold models tend to be affected by alignment errors 
in the input alignment.
To address this, averaged probability distributions of RNA sequences 
involved in the input alignment were introduced by Kiryu {\em et al.} \cite{pmid17182698}.
This leads to an MEG estimator with the probability distribution
%
\begin{align}
  p^{(\mbox{\scriptsize ave})}(\theta|A)=\frac{1}{n} \sum_{x\in A} p(\theta|x)
\end{align}
%
where $p(\theta|x)$ is a probabilistic model for RNA secondary structures, 
for example, the McCaskill model \cite{pmid1695107}, the CONTRAfold model \cite{pmid16873527}, the BL model \cite{pmid20940338,pmid19933322,pmid17646296}, and others \cite{DBLP:journals/jbcb/SatoHMAS10}.
Note that neither covariation nor phylogenetic information about alignments is considered in
this probability distribution.

In \cite{pmid17182698}, the authors utilized the McCaskill model as a probabilistic model
for individual RNA sequences (i.e., they used $p(\theta|x)$ in Eq.(\ref{eq:ave_p})), 
and showed that their method was more robust
with respect to alignment errors than RNAalifold and Pfold.
Using averaged probability distributions for RNA sequences in an input alignment 
is compatible with
Evaluation Procedure~\ref{def:eval_proc}. 
See Hamada {\it et al.}~\cite{pmid20843778} for a detailed discussion.

\subsubsection{Mixture of several distributions}\label{sec:mix_dist}

Hamada {\it et al.} \cite{pmid20843778} pointed out that arbitrary information 
can be incorporated into common secondary structure predictions 
by utilizing a {\em mixture} of probability distributions.
For example, the probability distribution
%
\begin{align}
  p(\theta|A)=
  w_1 \cdot p^{(\mbox{\scriptsize pfold})}(\theta|A) + 
  w_2 \cdot p^{(\mbox{\scriptsize alifold})}(\theta|A) + 
  w_3 \cdot p^{(\mbox{\scriptsize ave})}(\theta|A),\label{eq:mix_dist}
\end{align}
%
where $w_1$, $w_2$, and  $w_3$ are positive values that satisfy $w_1 + w_2 + w_3 = 1$,
includes 
covariation (Section~\ref{sec:CV}), phylogenetic tree (Section~\ref{sec:PI}), 
and majority rule (Section~\ref{sec:MR}) information.

\begin{figure}[ht]
\centerline{
\includegraphics[width=0.35\textwidth]{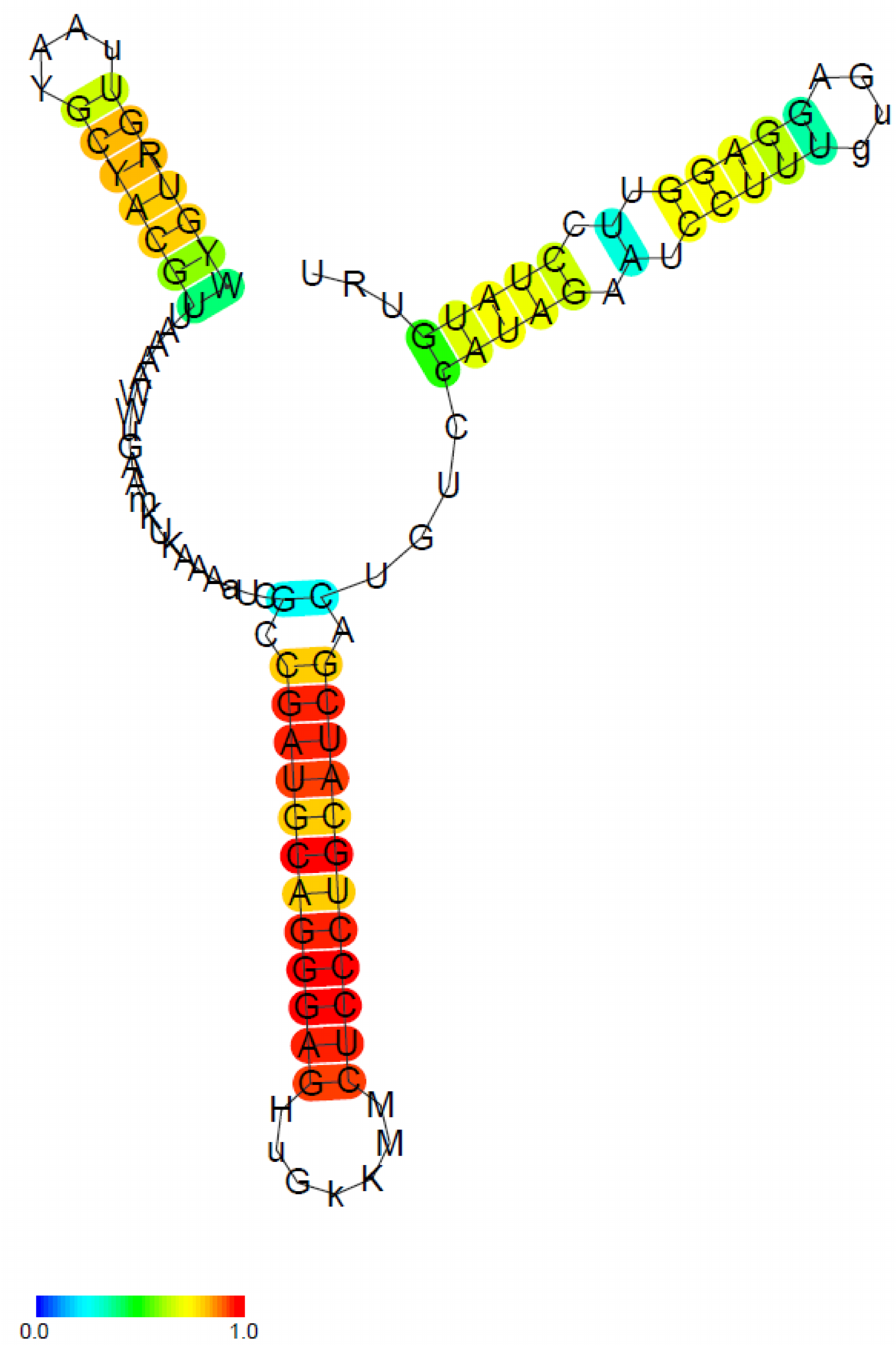}
}
\caption{\label{fig:centroidalifold_web}\small
Example of a predicted common secondary structure 
from the CentroidAlifold Web Server \cite{pmid19435882}
({http://www.ncrna.org/centroidfold}). 
The input is a multiple sequence alignment of the traJ $5'$ UTR.
The color of a base-pair indicates the averaged base-pairing probabilities 
(among RNA sequences in the alignment) of 
the base-pair, where warmer colors represent higher probabilities.
}
\end{figure}

In CentroidAlifold \cite{pmid20843778}, users can employ a mixed
distribution given by an arbitrary combination of
RNAalipffold (Section~\ref{sec:rnaalipffold_model}), 
Pfold (Section~\ref{sec:pfold-model}), and 
an averaged probability distribution (Section~\ref{sec:ave-dist}) 
based on the McCaskill or CONTRAfold model.
An example of a result from the CentroidAlifold Web Server is shown in Figure~\ref{fig:centroidalifold_web}, in
which colors of base-pairs indicate the marginal base-pairing probability with respect to this mixture distribution.
Hamada {\it et al.}  also showed that computation of MEG estimators 
with a mixture distribution can be easily conducted by utilizing base-pairing 
probability matrices (see also the next section), when certain gain functions are
employed.
Moreover, computational experiments indicated that 
CentroidAlifold with a mixture model is significantly better than RNAalifold, 
Pfold or McCaskill-MEA.

\subsection{Choice of gain functions $G(\theta,y)$}\label{sec:choice_gain}
%
A choice of the gain function in MEG estimators corresponds to the decoding method\footnote{%
Prediction of one final common secondary structure from the distribution.
}
of common RNA secondary structures,
given a probabilistic model of common secondary structures.

\subsubsection{The Kronecker delta function}\label{sec:delta-gain}
%
A straightforward choice of the gain function is the Kronecker delta function:
\begin{align}
  G^{(\mbox{\scriptsize delta})}(\theta,y)\left(=\delta(\theta,y)\right)=
  \left\{\begin{array}{ll}
  1 & \mbox{if $y$ is the exactly same as $\theta$}\\
  0 & \mbox{otherwise}
  \end{array}
  \right.\label{eq:delta_function}
\end{align}
The MEG estimator with this gain function  is equivalent to the 
`maximum likelihood estimator' (ML estimator) with respect to a given 
probabilistic model for RNA common secondary structures (cf. Section~\ref{sec:choice_prob}), which predicts
the secondary structure with the highest probability.

%
\subsubsection{The $\gamma$-centroid-type gain function \cite{pmid19095700}}\label{sec:gcent-gain} 
%
It is known that the probability of the ML estimation is extremely small, due to 
the immense number of secondary structures that could be predicted; this fact
is known as the `uncertainty' of the solution, 
and often leads to issues in bioinformatics \cite{hamada_2013}. 
Because the MEG estimator with the delta function considers only the solution 
with the highest probability, it is affected by this uncertainty.
A choice of gain function that partially overcomes this uncertainty of solutions 
is the $\gamma$-centroid-type gain function \cite{pmid19095700}:
\begin{align}
  G_\gamma^{(\mbox{\scriptsize centroid})}(\theta,y)=
  \sum_{i,j} \left[\gamma I(\theta_{ij}=1)I(y_{ij}=1) + I(\theta_{ij}=0)I(y_{ij}=0) \right]\label{eq:g-centroid_gain}
\end{align}
where $\gamma>0$ is a parameter that adjusts the relative importance of the SEN and PPV of base-pairs in a predicted structure 
(i.e., a larger $\gamma$ produces more base-pairs in a predicted secondary structure).
This gain function is motivated by the concept that 
more true base-pairs (TP and TN) and
fewer false base-pairs (FP and FN) should be predicted \cite{pmid19095700} when the
entire distribution of secondary structures is considered.
Note that, when $\gamma=1$, the gain function is equivalent to that of the centroid 
estimator \cite{pmid18305160}.

\subsubsection{The CONTRAfold-type gain function \cite{pmid16873527}}\label{sec:contra-gain}
%
In conventional RNA secondary structure predictions, 
another gain function has been proposed\footnote{%
  Historically, the CONTRAfold-type gain function was proposed earlier than
  the $\gamma$-centroid-type gain function. 
}, which is based on the number of accurate predictions
of every single position ({\em not} base-pair) in an RNA sequence.\footnote{%
This is not consistent with Evaluation Procedure~\ref{def:eval_proc}, because 
accurate predictions of {\em base-pairs} with respect to reference structures 
are evaluated in it.} The CONTRAfold-type gain function is
\begin{align}
  G_{\gamma}^{(\mbox{\scriptsize contra})}(\theta,y)
  =\sum_{i=1}^{|A|} 
  \Biggl[ 
    {\gamma}\sum_{j:j\ne i}  I(\theta_{ij}^*=1)I(y_{ij}^*=1)+ 
    \prod_{j:j\ne i} I(\theta_{ij}^*=0)I(y_{ij}^*=0)
    \Biggl]\label{eq:contra_gain}
\end{align}
where $\theta^*$ and $y^*$ are symmetric extensions of $\theta$ and $y$, respectively
(i.e., $\theta^*_{ij}=\theta_{ij}$ for $i<j$ and $\theta^*_{ij}=\theta_{ji}$ for $j<i$).
This gain function is also applicable to MEG estimators for common secondary structure predictions.
It should be emphasized that MEG estimators based on the CONTRAfold-type gain function
(for any $\gamma>0$) do not include centroid estimators, while the $\gamma$-centroid-type
gain function includes the centroid estimator as a special case (i.e. $\gamma=1$).

\subsubsection{Remarks about choice of gain function}
%
From a theoretical viewpoint, 
the $\gamma$-centroid-type gain function is more appropriate 
for Evaluation~Procedure~\ref{def:eval_proc} than either the delta function 
or the CONTRAfold-type gain function\footnote{%
The CONTRAfold-type gain function has a bias toward accurate predictions of 
base-pairs, compared to the $\gamma$-centroid-type gain function.}. which
is also supported by several empirical (computational) experiments.
See Hamada {\it et al.} \cite{pmid20843778} for a detailed discussion.

Another choice of the gain function is MCC (or F-score), which 
takes a balance between SEN and PPV of base-pairs, and the MEG estimator 
with 
this gain function leads to an algorithm that maximizes 
{\em pseudo}-expected accuracy \cite{pmid21118522}.
In addition, the estimator with this gain function 
includes only one parameter 
for predicting secondary 
structure\footnote{The $\gamma$-centroid-type and CONTRAfold-type 
gain functions contain a parameter adjusting the ratio of SEN and PPV 
for a predicted secondary structure.}.


\subsection{Computation of common secondary structure through MEG estimators}
%
\subsubsection{MEG estimator with delta function}
The MEG estimator with the delta function 
(that predicts the common secondary structure with the highest probability with respect to
a given probabilistic model) 
can be computed by employing a CYK (Cocke-Younger-Kasami)-type algorithm.
For example, see \cite{pmid15313604} for details.

\subsubsection{MEG estimator with $\gamma$-centroid (or CONTRAfold) type gain function}
The MEG estimator with the $\gamma$-centroid-type 
(or CONTRAfold-type) gain function is computed based on  
`base-pairing probability matrices' (BPPMs)\footnote{%
The BPPM is a probability matrix $\{p_{ij}\}$
in which $p_{ij}$ is the marginal probability that 
the $i$-the base $x_i$ and the $j$-th base $x_j$ form a base-pair
with respect to a given probabilistic distribution of secondary structures.
For many probabilistic models,
including the McCaskill model and the CONTRAfold model, 
the BPPM for a given sequence can
be computed efficiently by utilizing inside-outside algorithms. 
See \cite{pmid1695107} for the details.} and 
Nussinov-style dynamic programming (DP) \cite{pmid20843778,pmid16873527}:
\begin{equation}
M_{i,j} = \max \left \{
\begin{array}{ll}
M_{i+1,j} \\
M_{i,j-1} \\
M_{i+1,j-1} + S_{ij}\\ 
\max_k \left[M_{i,k} + M_{k+1,j}\right] 
\end{array}
\right.\label{eq:dp_centroidalifold}
\end{equation}
where 
$M_{i,j}$ is the optimal score of the subsequence $x_{i\cdots j}$ and
$S_{ij}$ is a score computed from the BPPM(s).
For instance, for the $\gamma$-centroid estimator with the RNAalipffold model,
the score $S_{ij}$ is equal to $S_{ij}=(\gamma+1)p_{ij}^{(\mbox{\scriptsize alipffold})}-1$
where 
$p_{ij}^{(\mbox{\scriptsize alipffold})}$ is the base-pairing probability with respect to
the RNAalipffold model.
This DP algorithm maximizes the sum 
of (base-pairing) probabilities 
$p_{ij}^{(\mbox{\scriptsize alipffold})}$  which are larger than $1/(\gamma+1)$,
and requires $O(|A|^3)$ time.

\subsubsection{MEG estimators with average probability distributions}
The MEG estimator with an average probability distribution (Section~\ref{sec:ave-dist})
can be computed by using
averaged base-pairing probabilities, $\{p_{\ij}\}_{i<j}$:
\begin{align}
  p^{\mbox{\scriptsize(ave)}}_{ij}=\frac{1}{n} \sum_{x\in A} p^{(x)}_{ij} 
\end{align}
where
\begin{align}
  p_{ij}^{(x)}=
  \left\{
  \begin{array}{ll}
    \sum_{\theta\in\mathcal{S}(x)}I(\theta_{\tau(i)\tau(j)}=1)p(\theta|x')& \mbox{if both $x_i$ and $x_j$ are not gaps}\\
    0 & \mbox{otherwise}
  \end{array}\right.\label{eq:ave_p}
\end{align}
In the above,
$x'$ is the RNA sequence given by removing gaps from $x$ and $n$ is the number of 
sequences in the alignment $A$.
The function $\tau(i)$ returns the position in $x'$ corresponding to the position 
$i$ in $x$.

The common secondary structure of MEG estimator with the $\gamma$-centroid gain function
And the average probability distribution are computed by using the DP recursion
in  Eq.(\ref{eq:dp_centroidalifold}) where $S_{ij}=(\gamma+1)p^{(\mbox{\scriptsize ave})}_{ij}-1$. This procedure has a time complexity of $O(n|A|^3)$ where
$n$ is the number of sequences in the alignment.

\subsubsection{MEG estimators with a mixture distribution}

The MEG estimator with a mixture of distribution 
(Section~\ref{sec:mix_dist}) and the delta function (Section~\ref{sec:delta-gain}) 
cannot be computed efficiently.
However, if the $\gamma$-centroid-type (or CONTRAfold-type) gain function 
is utilized, the prediction can be conducted using a similar DP recursion to that in 
Eq.~(\ref{eq:dp_centroidalifold}).
For instance, the DP recursion of the  $\gamma$-centroid-type gain function
with respect to Eq.~(\ref{eq:mix_dist}) is equivalent to the one 
in Eq.(\ref{eq:dp_centroidalifold}) 
with $S_{ij} = (\gamma+1) p_{ij}^* - 1$
where
\begin{align}
p_{ij}^*=
w_1 \cdot p_{ij}^{(\mbox{\scriptsize pfold})} + 
w_2 \cdot p_{ij}^{(\mbox{\scriptsize alipffold})} + 
\frac{w_3}{n}\sum_{x\in A}p^{(x)}_{ij}.\label{eq:pij_centroid}
\end{align}
In the above, $p_{ij}^{(\mbox{\scriptsize pfold})}$ and $p_{ij}^{(\mbox{\scriptsize alipffold})}$ 
are base-pairing probabilities
for the Pfold and RNAalipffold models, respectively, and
$\{p_{ij}^{(x)}\}$ is a base-pairing probability matrix with respect to 
a probabilistic model for secondary structures of single RNA sequence $x$ 
(McCaskill or CONTRAfold model).
%
Note that the total computational time of CentroidAlifold with a mixture of 
distributions still remains $O(n|A|^3)$.

\subsubsection{MEG estimators with probability distribution including pseudoknots}
Using probability distributions of secondary structures 
with pseudoknots in MEG estimators generally has higher 
computational cost \cite{pmid23057823}.
To overcome this, for example, 
IPKnot \cite{pmid21685106} utilizes an approximated method for 
determining the probability distribution 
as along with integer linear programming  for predicting a 
final common secondary structure.

\section{A classification of tools for Problem~\ref{prob:rnac}}\label{sec:tools}

Table~\ref{tab:list_of_tools}, shows a comprehensive list of tools 
for common secondary structure prediction from aligned RNA sequences
(in alphabetical order within groups that do or do not consider pseudoknots 
\footnote{%
Tools for predicting common secondary structures {\em without} 
pseudoknots are much faster than those for predicting secondary structures {\em with} pseudoknots.%
}%
). 
To the best of my knowledge, Table~\ref{tab:list_of_tools} is a complete list 
of tools for Problem~\ref{prob:rnac} as of 17 June 2013.

In Table~\ref{tab:comparison},
the tools in Table~\ref{tab:list_of_tools} are classified
based on the considerations of Sections~\ref{sec:info} and~\ref{sec:classification}.
The classification leads to much useful information: 
(i) the pros and cons of each tool; (ii) the similarity (or dissimilarity) 
among tools; (iii) which tools are more suited to 
Evaluation Procedure~\ref{def:eval_proc}; and 
(iv) a unified framework within which to design algorithms for Problem~\ref{prob:rnac}.
I believe that the classification will bring a deeper understanding 
of each tool, although several tools 
(which are not based on probabilistic models and 
depend fundamentally on heuristic approaches) cannot be classified 
in terms of MEG estimators.

\begin{table}[th]
  \caption{\label{tab:comparison}
    Comparison of tools (in Table~\ref{tab:list_of_tools}) for common secondary structure predictions 
    from aligned sequences
  }
  \centerline{\small
    {\small
\begin{tabularx}{\textwidth}{lccccccccclX}
\toprule
     &     \multicolumn{2}{c}{Software$^{\mathrm{a}}$} & \multicolumn{7}{c}{Used information$^{\mathrm{b}}$} & Gain$^{\mathrm{c}}$ & Prob.dist.$^{\mathrm{d}}$\\
\cmidrule{4-10}
Name & SA & WS                           & TS & ML & CV & PI & MI & MR & EI\\
\midrule
\multicolumn{4}{l}{\bf (Without pseudoknot)}  \\
CentroidAlifold & \checkmark & \checkmark & \checkmark & \checkmark & \checkmark & \checkmark &  & \checkmark & {\bf\scriptsize(*1)} & $\gamma$-cent & Any {\bf\scriptsize(*2)} \\ 
ConStruct & \checkmark &  & \checkmark &  & \checkmark &  & \checkmark & \checkmark &  & na & na\\
KNetFold & & \checkmark & \checkmark &  & & & \checkmark & & & na & na\\
McCaskill-MEA  & \checkmark &  &  &  &  &  &  & \checkmark & & contra & Av(Mc)\\
PETfold  & \checkmark & \checkmark &  &  &  & \checkmark &  &  &  & contra & {\scriptsize Pf+Av(Mc)}\\
Pfold  & \checkmark &  &  &\checkmark  &  & \checkmark &  & & & delta & Pf\\
PhyloRNAalifold  & \checkmark & & & & \checkmark & \checkmark & & & & delta & Ra\\
PPfold  & \checkmark &  &  & \checkmark &  & \checkmark &  &  & \checkmark & delta & Pf\\
RNAalifold  & \checkmark & \checkmark & \checkmark &  & \checkmark &  &  &  & & delta & Ra\\
RSpredict  & \checkmark & \checkmark &  \checkmark &  & \checkmark &  &  &  & & na & na\\ 
\multicolumn{4}{l}{\bf (With pseudoknot)}  \\
hxmatch  & \checkmark  &  & \checkmark &  & \checkmark &  &  & & & na & na\\
ILM & \checkmark  & \checkmark & \checkmark &  &  &  & \checkmark &  &  & na & na\\
IPKnot & \checkmark & \checkmark & \checkmark & \checkmark & \checkmark &  &  & \checkmark & & $\gamma$-cent & Any {\bf\scriptsize(*2)}\\
MIfold  & \checkmark {\bf \scriptsize(*3)} &  &  &  &  &  & \checkmark &  & & na & na\\
\bottomrule
\end{tabularx}
}

  }
             {\footnotesize
               $^{\mathrm{a}}$ Type of  software available. 
               SA: stand alone; WS: Web server, 
               TS: Thermodynamic stability (Section~\ref{sec:TS}).\\
               $^{\mathrm{b}}$
               In the `Information used' columns,
               ML: Machine learning (Section~\ref{sec:ML});
               CV: Covariation (Section~\ref{sec:CV}); 
               PI: Phylogenetic (evolutionary) information (Section~\ref{sec:PI});
               MI: Mutual information (Section~\ref{sec:MI});
               MR: Majority rule (Section~\ref{sec:MR});
               EI: Experimental information (Section~\ref{sec:EI});\\
               $^{\mathrm{c}}$ In the column `Gain',
               $\gamma$-cent: $\gamma$-centroid-type gain function (Section~\ref{sec:gcent-gain});
               contra: CONTRAfold-type gain function (Section~\ref{sec:contra-gain}).\\
               $^{\mathrm{d}}$~In the column `Prob.dist',
               Pf: Pfold model (Section~\ref{sec:pfold-model});
               Ra: RNAalipffold model (Section~\ref{sec:rnaalipffold_model});
               Av(Mc): Averaged probability distribution with McCaskill model (Section~\ref{sec:ave-dist});
               `+' indicates a mixture distribution (of several models).
               `na' means `Not available' due to no use of probabilistic models.\\
               {\bf\scriptsize(*1)} If the method proposed in \cite{pmid23210474} is used, 
               experimental information
               derived from SHAPE \cite{pmid20554050} and PARS \cite{pmid20811459} 
               is easily incorporated in CentroidAlign.\\
               {\bf\scriptsize(*2)} CentroidAlifold can employ a mixed
               distribution given by an {\em arbitrary} combination of
               RNAalipffold, Pfold, and an averaged probability distribution based on the 
               McCaskill or CONTRAfold models.\\
               {\bf\scriptsize(*2)} MATLAB codes are available.
             }
\end{table}

\section{Discussion}\label{sec:discussion}

\subsection{Multiple sequence alignment of RNA sequences}

Predicting a multiple sequence alignment (point estimation) 
from unaligned sequences 
is not reliable because the probability of the alignment becomes extremely small.
This is called the `uncertainty' of alignments which raises serious 
issues in bioinformatics \cite{hamada_2013}.
In one {\em Science} paper \cite{pmid18218900}, for instance, the authors argued 
that the uncertainty of multiple sequence
alignment greatly influences phylogenetic topology estimations: phylogenetic topologies
estimated from multiple alignments predicted by five widely used aligners are different from
one another.
Similarly,
point estimation of multiple sequence alignment will greatly affect 
consensus secondary structure prediction. 

In Problem~\ref{prob:rnac},
because the quality of the multiple sequence alignment 
influences the prediction of common secondary structure,
the input multiple alignment should be given by a multiple aligner 
which is designed specifically for RNA sequences.
Although strict algorithms for multiple alignments taking into account secondary 
structures are equivalent to the Sankoff algorithm \cite{sankoff1985} and have
huge computational costs, several multiple aligners which are 
fast enough to align long RNA sequences are available: these are
CentroidAlign \cite{pmid23507751,pmid19808876}, R-coffee \cite{pmid18420654}, 
PicXXA-R \cite{pmid21342569}, DAFS \cite{pmid23060618}, and MAFFT \cite{pmid18439255}.
In those multiple aligners, not only nucleotide sequences but also secondary
structures are considered in the alignment, and they are, therefore, suitable for
generating input multiple alignments for Problem~\ref{prob:rnac}.

Because the common secondary structure depends on multiple alignment,
an approach adopted in
RNAG \cite{pmid21788211} also seems promising. This approach iteratively samples from the
conditional probability distributions P(Structure $|$ Alignment) and
P(Alignment $|$ Structure). 
Note, however, that RNAG does not solve Problem~\ref{prob:rnac} directly.

\subsection{Improvement of RNA secondary structure predictions using common secondary structure}

Although several studies have been conducted for RNA secondary structure 
predictions for a single RNA sequence~\cite{pmid23511969,pmid10329189,pmid19095700,pmid16873527}, 
the accuracy is still limited, especially for long RNA sequences.
By employing comparative approaches using homologous sequence information, 
the accuracy  of RNA secondary structure prediction will be improved.
In many cases, homologous RNA sequences of the target RNA sequence are obtained,
and someone would like to know the common secondary structure of those sequences.
Gardner and Giegerich \cite{pmid15458580} introduced 
three approaches for comparative analysis of RNA sequences, and
common secondary structure prediction is essentially utilized in the first of these.
However, 
if the aim is to improve the accuracy of secondary structure predictions,
common secondary structure prediction is not always the best solution,
because it is not designed to predict the optimal secondary structure 
of a specific target RNA sequence.
If you have a target RNA sequence for which the secondary structure is to be predicted,
the approach adopted by the CentroidHomfold \cite{pmid19478007,pmid21565800} software 
is more appropriate than a method based on 
common RNA secondary structure prediction.

\subsection{How to incorporate several pieces of information in algorithms}\label{sec:how_to_inc}

As shown in this review, there are two ways to incorporate several pieces of information into 
an algorithm for common secondary structure prediction.
The first approach is to modify the (internal) algorithm itself in order to handle 
the additional information.
For example, 
PhyloRNAalifold \cite{pmid23621982} incorporates phylogenetic information into
the RNAalifold algorithm by modifying the internal algorithm and
PPfold \cite{pmid22877864} modifies the Pfold algorithm to handle experimental 
information.
The drawbacks of this approach are the relatively large implementation cost
and the heuristic combination of the information.

On the other hand, another approach adopted in CentroidAlifold \cite{pmid20843778} 
is promising because it can easily incorporate 
many pieces of information into predictions if a base-pairing probability matrix is available.
Because the approach depends on only base-pairing probability matrices,
and does not depend on the detailed design of the algorithm, it is easy to implement
an algorithm using a mixture of distributions.

Moreover, a method to update a base-pairing probability matrix 
(computed using sequence information only)
which incorporates experimental information \cite{pmid23210474} has recently been proposed. 
The method is independent of the probabilistic models 
of RNA secondary structures,
and is suitable for incorporating experimental information 
into common RNA secondary structure prediction.
A more sophisticated method by Washietl {et al.} \cite{pmid22287623} can also
be used to incorporate experimental information into common secondary structure predictions, 
because it produces a BPPM that takes experimental information into account.

\subsection{A problem that is mathematically related to Problem~\ref{prob:rnac}}
%
Problem~\ref{prob:rnac}, which is considered in this paper, can be extended to predictions of RNA-RNA interactions, another important task 
in RNA bioinformatics (e.g., \cite{pmid20823308,pmid22767260}).
%
\begin{problem}[Common joint structure predictions 
of two aligned RNA sequences]\label{prob:rnarna}
Given two multiple alignments $A_1$ and $A_2$ of RNA sequences, then predict
a joint secondary structure between $A_1$ and $A_2$.
\end{problem}
%
Because the mathematical structure of Problem~\ref{prob:rnarna} 
is similar to that of Problem~\ref{prob:rnac},
the ideas utilized in designing the algorithms for 
common secondary structure predictions 
can be adopted in the development of methods for this new problem.
In fact, Seemann {\it et al.}  \cite{pmid21088024,pmid21609960} have employed similar idea, 
adapting the PETfold algorithm to Problem~\ref{prob:rnarna}
(implemented in the PETcofold software).
Note that the problem of (pairwise) alignment between two multiple {\em alignments} 
(cf. see \cite{pmid21365017} for the details) has
a similar mathematical structure to Problem~\ref{prob:rnac}.

\section{Conclusion}\label{sec:conclusion}

In this review, I focused on RNA secondary structure predictions 
from {\em aligned} RNA sequences, in which a secondary structure 
whose length is equal to the length of the input alignment is predicted.
A predicted common secondary structure is useful not only for further functional
analyses of the ncRNAs being studied but also for improving RNA secondary structure predictions and
for finding ncRNAs in genomes.
In this review, I systematically classified existing algorithms on the basis of 
(i) the information utilized in the algorithms
and 
(ii) the corresponding MEG estimators, which consist of a gain function 
and a probability distribution of common secondary structures. This classification 
will provide a deeper understanding of each algorithm. 

\section*{Acknowledgement}
%
This work was supported in part by MEXT KAKENHI 
(Grant-in-Aid for Young Scientists (A): 24680031; 
Grant-in-Aid for Scientific Research (A): 25240044).

\bibliographystyle{unsrt}{\small
  \bibliography{bib-paper,bib-mhamada}
}

\end{document}